\documentclass[12pt]{iopart}
\usepackage{graphicx}
\usepackage{hyperref}
\begin{document}

\title[Trilinear Hamiltonian: Modeling Hawking Radiation from a Quantized Source]{The Trilinear Hamiltonian: A Zero Dimensional Model of Hawking Radiation from a Quantized Source}

\author{Paul D. Nation and Miles P. Blencowe}
\address{Department of Physics and Astronomy, Dartmouth College, Hanover, New Hampshire 03755, USA}
\eads{paul.d.nation@dartmouth.edu}
\begin{abstract}
We investigate a quantum parametric amplifier with dynamical pump mode, viewed as a zero-dimensional model of Hawking radiation from an evaporating black hole.  The conditions are derived under which the spectrum of particles generated from vacuum fluctuations deviates from the thermal spectrum predicted for the conventional parametric amplifier.  We find that significant deviations arise when the pump mode (black hole) has emitted nearly half of its initial energy into the signal (Hawking radiation) and idler (in-falling particle) modes.  As a model of black hole dynamics, this finding lends support to the view that late-time Hawking radiation contains information about the quantum state of the black hole and is entangled with the black hole's quantum gravitational degrees of freedom. 
\end{abstract}

\pacs{04.70.Dy, 42.65.Lm, 03.67.Bg}
\submitto{New Journal of Physics focus issue ``Classical and Quantum Analogues for Gravitational Phenomena and Related Effects"}
\maketitle

\section{\label{sec:introduction}Introduction}
In the 35 years since Hawking's seminal paper on the quantum emission of radiation from a black hole\cite{hawking:1974}, a large body of work has been devoted to solving the so called information loss problem.  For a black hole of fixed mass $M$, this emission process yields a black body spectrum with characteristic temperature $T_{H}=\hbar c^{3}/ 8\pi k_{\mathrm{B}}GM$, irrespective of the initial state of the matter from which the black hole is formed.  The inability to reconstruct the initial, possibly pure state of the black hole from the total emitted radiation signals the apparent breakdown of unitary evolution and the $S$-matrix description of the Hawking process.  Although Hawking's calculations have since been verified in a number of ways\cite{hawking:1975,hartle:1976,boulware:1976,gibbons:1977,parentani:2000}, this breakdown at the foundation of quantum mechanics suggests that our understanding of black hole dynamics is not yet complete. 

The information loss problem rests on two key assumptions made in the standard derivation of Hawking radiation: (1) the perfectly thermal (i.e. mixed) character of the outgoing radiation; (2) the validity of this emission process over the lifetime of the black hole.  The notion of thermal spectrum considered here is not that of a blackbody frequency spectrum, but rather the quantum thermal probability distribution defined by the temperature $T_{H}$ of a single mode (single frequency) of Hawking radiation.  The traditional picture of the Hawking process leaves no room for deviations from this thermal distribution and thus breaks the requirement of pure-state$\rightarrow$pure-state evolution enforced by unitarity.  By itself, this process need not lead to information loss since the information content of the black hole may be stored in entanglement between particle pairs created on opposite sides of the horizon\cite{bombelli:1986,srednicki:1993}.  Results from many-body theory suggesting that entanglement across a boundary scales with the area of the boundary lends credence to this view\cite{eisert:2010}.  However, with the second assumption, the black hole causes a loss of entanglement and thus information.

With the expectation that information must be conserved, many suggestions for resolving the information loss problem have been put forward.  Current proposed solutions include long-lived and stable Planck-scale remnants\cite{aharanov:1987,giddings:1992}, baby universes\cite{hawking:1988,frolov:1990}, and the possibility of information escaping as non-thermal Hawking radiation \cite{page:1993,parikh:2000}.  In all of these proposals, corrections to the Hawking process manifest themselves when the black hole has evaporated to a size near that of the Planck length, $l_{p}=\sqrt{\hbar G/c^{3}}\approx 10^{-35}~\mathrm{m}$, at which quantum gravitational effects, neglected in Hawking's original analysis, are expected to play a role.  In considering quantum states of the gravitational field, there is the possibility of back-reaction and entanglement of the radiating matter degrees of freedom with those of gravity\cite{hawking:2005,terno:2005}.  Although it is natural to consider a quantized gravitational field for the Hawking process, the current lack of a full quantum mechanical description of gravity severely limits progress in directly addressing this scenario.  In fact, exactly which degrees of freedom, if any, should be quantized is still subject to debate\cite{jacobson:1995,carlip:2008}.  In this paper we investigate a simple, zero-dimensional quantum optics model of Hawking radiation that mimics some of the essential physics present in the original information loss problem.

\begin{figure}[htbp]
\begin{center}
\includegraphics[width=4.0in]{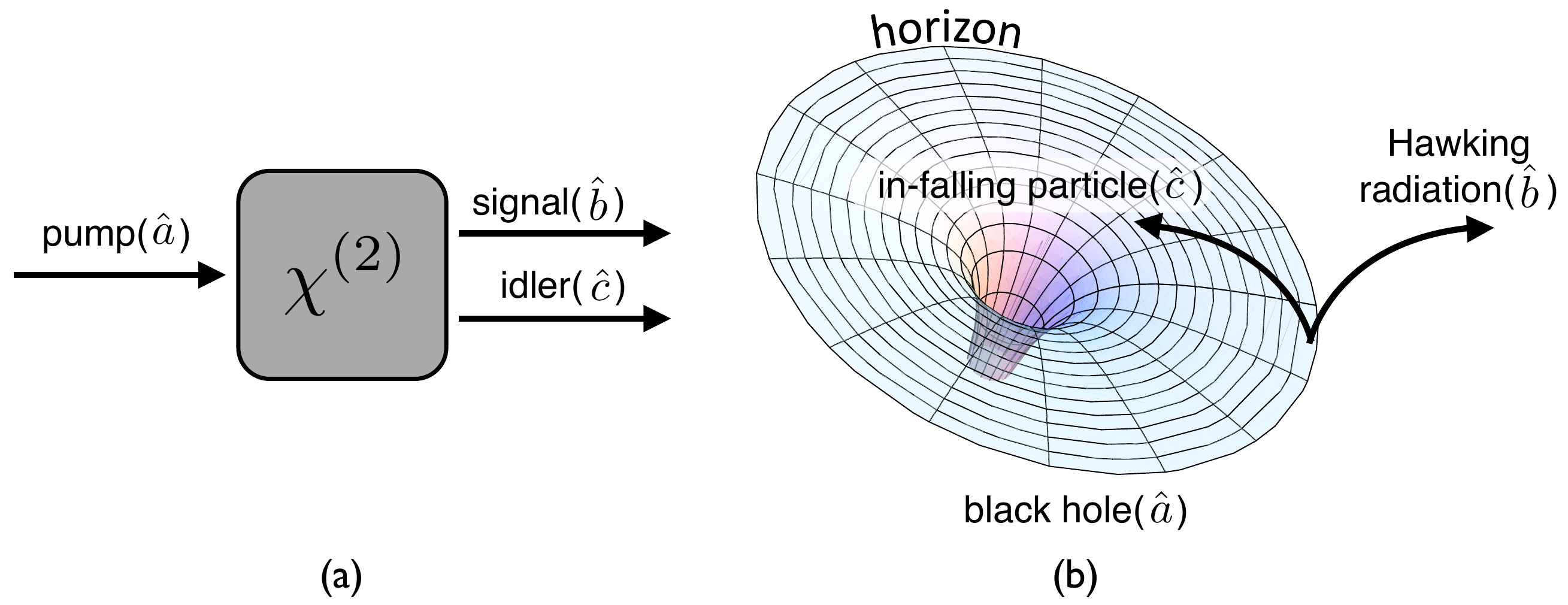}
\caption{a.)Diagram depicting the dynamics of the trilinear Hamiltonian.  Initial vacuum modes are not shown for clarity.  The nonlinear interaction is generated by a system possessing a second-order susceptibility $\chi^{(2)}$.  b.)Equivalent dynamical elements involved in the Hawking process.}
\label{fig:relation}
\end{center}
\end{figure}

As a zero-dimensional model we consider the following trilinear Hamiltonian:
\begin{equation}\label{eq:trilinear}
\hat{H}=\hbar\omega_{a}\hat{a}^{+}\hat{a}+\hbar\omega_{b}\hat{b}^{+}\hat{b}+\hbar\omega_{c}\hat{c}^{+}\hat{c}+i\hbar\chi\left(\hat{a}\hat{b}^{+}\hat{c}^{+}-\hat{a}^{+}\hat{b}\hat{c}\right),
\end{equation}
consisting of three harmonic oscillator modes with the frequency relation, $\omega_{a}=\omega_{b}+\omega_{c}$.  We will designate the modes as pump ($\hat{a}$), signal ($\hat{b}$), and idler ($\hat{c}$) respectively. This Hamiltonian describes several quantum optics processes including frequency conversion, Raman and Brillouin scattering, the interaction of two-level atoms with a single mode resonant EM field, and is the full quantum generalization of the parametric amplifier\cite{dicke:1953,mollow:1967,travis:1968,tucker:1969,walls:1970,lu:1973,agrawal:1974,mcneil:1983}.  The connection between black hole radiance and  parametric amplification was appreciated shortly after Hawking's discovery\cite{gerlach:1976}. Both processes amplify vacuum fluctuations resulting in the production of correlated photon pairs.  Tracing over one of the two subsystems (i.e., signal and idler) yields statistics that are identical to that of a thermal distribution\cite{barnett:1985,yurke:1987}.  The energy source in the parametric amplifier is assumed to be a classical pump such as a laser or microwave generator with fixed amplitude driving a system with $\chi^{(2)}$ nonlinearity, the first nonlinear susceptibility in a medium without inversion symmetry\cite{boyd:2003}.  Viewed as a black hole model, the pump plays the role of black hole mass $M$ while the signal and idler modes of the parametric amplifier represent the escaping and trapped Hawking photons, respectively, as depicted schematically in Fig.~\ref{fig:relation}.
The trilinear Hamiltonian (\ref{eq:trilinear}) generalizes the parametric amplifier by quantizing the pump mode and allowing for energy loss to the signal and idler modes.  The expectation value of the pump mode energy is analogous to the mass $M$ of a quantum mechanical black hole.  We will explore this model by establishing the conditions under which the signal mode spectrum of the trilinear Hamiltonian deviates from the predicted  thermal spectrum of the conventional parametric amplifier.  We shall see that the quantization of the pump mode degree of freedom results over time in entanglement with the signal and idler modes and dynamics that become markedly different from the parametric approximation. In particular, the signal mode develops a strongly non-thermal spectrum that is dependent on the initial pump mode state. The corresponding entropy is reduced relative to that of a thermal (maximally mixed) state indicating the presence of information.  These model system results lend support to the view that late-time Hawking radiation contains an increasing amount of information about the initial quantum state of the black hole and is composed of particles entangled with quantized gravitational states. 

The outline of this paper is as follows:  In Sec.~\ref{sec:parametric} we review the derivation of the amplification of vacuum fluctuations under the parametric approximation.  Sec.~\ref{sec:semiclassical} considers the semi-classical approximation whereby the back-reaction from the quantized radiation onto the classical pump is accounted for and derives the self-consistent equations of motion for the signal and idler modes.  In Sec.~\ref{sec:quantum} we consider the full quantum dynamics Eq.~(\ref{eq:trilinear}) under the short time analytical approximation and compare this result along with the full quantum numerical solution to the parametric and semi-classical approximations from the previous sections.  Sec.~\ref{sec:entanglement} investigates the role of entropy and entanglement in the production of non-thermal states.  Finally, Sec.~\ref{sec:conclusion} concludes with a brief discussion of the results and consequences for black hole evaporation.

\section{Parametric Amplifier and Hawking Emission}\label{sec:parametric}
In the following,  we  derive the well-known thermal spectrum of the signal mode under the parametric assumption of a fixed amplitude pump mode.  Replacing the pump mode in Eq.~(\ref{eq:trilinear}) with a fixed amplitude drive $A$ results in the interaction frame Hamiltonian
\begin{equation}\label{eq:parametric}
H_{I}=i\hbar\chi A\left(b^{+}c^{+}-bc\right).
\end{equation}
The Heisenberg equations of motion for the signal and idler mode operators are,
\begin{equation}
\frac{db(t)}{dt}=A\chi c(t)^{+};\qquad \frac{dc(t)}{dt}=A\chi b(t)^{+}.
\end{equation}
These can be readily solved yielding the Bogoliubov transformations,
\begin{equation}\label{eq:semi-motion}
\eqalign{
b(\tau)=b(0)\cosh\left(A\tau\right)+c(0)^{+}\sinh\left(A\tau\right) \\
c(\tau)=c(0)\cosh\left(A\tau\right)+b(0)^{+}\sinh\left(A\tau\right),
}
\end{equation}
where we have expressed the dynamics in dimensionless time $\tau=\chi t$.   If the system starts with both signal and idler modes in the ground state, $\left|\Psi(0)\right>=\left| 0,0\right>_{bc}$, then Eq.~(\ref{eq:semi-motion}) gives for the number operators $N_{b}$ and $N_{c}$:
\begin{equation}\label{eq:occupation}
N_{b}(\tau)=N_{c}(\tau)=\sinh^{2}\left(A\tau\right).
\end{equation}
Additionally, we are interested in the probability distribution of the individual signal and idler subsystems in the number state basis.  With the system initially in the ground state, the unitary evolution corresponding to Eq.~(\ref{eq:parametric}) can be expressed as,
\begin{equation}
\left| \Psi(\tau)\right>=\exp\left[A\tau\left(b^{+}c^{+}-bc\right)\right]\left| 0,0\right>_{bc}.
\end{equation}
Making use of the disentangling theorem\cite{truax:1985},
\begin{equation}\label{eq:disentangle}
\exp\left[A\tau\left(b^{+}c^{+}-bc\right)\right]=e^{\Gamma b^{+}c^{+}}e^{-g\left(b^{+}b+c^{+}c+1\right)}e^{-\Gamma bc},
\end{equation} 
with $\Gamma=\tanh\left(A\tau\right)$ and $g=\ln\cosh\left(A\tau\right)$,  where the last term in Eq.~(\ref{eq:disentangle}) vanishes for the ground state and the middle term reduces to $e^{-g}$, we have
\begin{equation}
e^{\Gamma b^{+}c^{+}}\left| 0,0\right>_{bc}=\sum_{n=0}^{\infty}\frac{\Gamma^{n}\left(b^{+}c^{+}\right)^{n}}{n!}\left| 0,0\right>_{bc}=\sum_{n=0}^{\infty}\tanh^{n}\left(A\tau\right)\left| n,n\right>_{bc},
\end{equation}
with the evolving state vector $\left| \Psi(\tau)\right>$ given by 
\begin{equation}\label{eq:squeezed}
\left| \Psi(\tau)\right>=\mathrm{sech}\left(A\tau\right)\sum_{n}\tanh^{n}\left(A\tau\right)\left| n,n\right>_{bc}.
\end{equation}
Let us now focus on operators acting on a subsystem spanned by the states of either the signal or idler mode individually.  Assuming we are interested in the signal mode only, tracing over the idler subsystem in Eq.~(\ref{eq:squeezed}) gives the operator expectation value
\begin{equation}\label{eq:semi-spectrum}
\left< O(\tau)\right>_{b}=\mathrm{sech}^{2}\left(A\tau\right)\sum_{n}\tanh^{2n}\left(A\tau\right)\left< n|O| n\right>_{b}
\end{equation}
for an arbitrary signal mode operator $O$.  Comparing Eq.~(\ref{eq:semi-spectrum}) with the spectrum of a thermal state defined by temperature $T$,
\begin{equation}\label{eq:semi-thermal}
\left< O\right>=\sum_{n}P_{n}\left< n_{b}|O|n_{b}\right>=\left(1-e^{-\hbar\omega_{b}/k_{\mathrm{B}}T}\right)\sum_{n}e^{-n\hbar\omega_{b}/k_{\mathrm{B}}T}\left< n_{b}|O|n_{b}\right>,
\end{equation}
indicates that the signal mode is in a thermal state provided we define the temperature as
\begin{equation}\label{eq:temperature}
T\left(\tau\right)=\frac{\hbar\omega_{b}}{2k_{\mathrm{B}}\ln\left[\coth\left(A\tau\right)\right]},
\end{equation}
where the time dependence is a consequence of the rapid increase in the occupation number of the modes, Eq.(\ref{eq:occupation}).  As in the Hawking process\cite{fuentes:2005}, the particle pairs generated by the parametric amplifier form an entangled two-mode squeezed state given by Eq.~(\ref{eq:squeezed})\cite{walls:2008}.  The bipartite structure of this system allows calculating the entanglement between particle pairs via the Von-Neumann entropy $S_{i}$, also referred to as entanglement entropy. For mode $i=b,c$,  using the reduced density matrix $\rho_{i}$ we have
\begin{equation}\label{eq:VNentropy}
S_{i}=-\mathrm{Tr}\left(\rho_{i}\ln\rho_{i}\right),
\end{equation}
where we drop the usual $k_{\mathrm{B}}$ factor.  For both the signal and idler modes, this entropy is given by
\begin{equation}\label{eq:thermal-entropy}
S^{\mathrm{th}}=-\ln\left[1-e^{-\hbar\omega/k_{\mathrm{B}}T(\tau)}\right]-\frac{\hbar\omega}{k_{\mathrm{B}}T(\tau)}\left[1-e^{\hbar\omega/k_{\mathrm{B}}T(\tau)}\right]^{-1},
\end{equation}
which is the thermal entropy of a quantum harmonic oscillator with temperature defined by Eq.~(\ref{eq:temperature}).  Thus, as for the Hawking thermal radiation,  we see that the temperature of  the parametric oscillator is determined by the entropy generated from tracing over one of the two modes in a particle pair squeezed state. 

\section{Semi-Classical Analysis: Backreaction on a Classical Pump Mode}\label{sec:semiclassical}
We now go beyond the just-considered parametric amplifier model and incorporate backreaction effects due to the emission process using a semi-classical approximation where the pump mode is treated as a classical variable, while the signal and idler modes are quantized.  In order to self consistently solve for the system evolution, we first work out the dynamics of the pump mode assuming it behaves as a classical variable affected by the expectation values of the signal and idler modes, after which we substitute the resulting c-number expressions for the pump operators into the Hamiltonian and calculate the signal and idler evolution.  A similar procedure arises for the semi-classical Einstein equations:
\begin{equation}
G_{\mu\nu}=\frac{8\pi G}{c^{4}}\left<\Psi |\hat{T}_{\mu\nu}|\Psi\right>,
\end{equation}
where the mass $M$ black hole spacetime geometry is modeled classically through the usual Einstein tensor $G_{\mu\nu}$,  interacting with the quantum matter/radiation fields via the expectation value of the stress-energy-momentum tensor operator $\hat{T}_{\mu\nu}$ with respect to a suitable incoming quantum field state $\left| \Psi\right>$.  Solving these semiclassical equations using an analogous procedure to that outlined above results in a nonlinear Schr\"{o}dinger equation\cite{kibble:1980} and also a possible loss of spacetime stability\cite{horowitz:1980}. In addition, issues such as non-causal dynamics\cite{anselmi:2007} and  possible inconsistencies with wavefunction collapse\cite{unruh:1984} occur.  In contrast, the semi-classical model presented here can be readily solved by exploiting the symmetries present in the Hamiltonian.  

To begin, we consider Eq.~(\ref{eq:trilinear}) in the interaction frame,
\begin{equation}\label{eq:interaction}
H_{I}=i\hbar\chi\left(ab^{+}c^{+}-a^{+}bc\right)
\end{equation}
and obtain the following mode equations,
\begin{equation}\label{eq:semiQ_dt}
\frac{da}{dt}=-\chi bc;\qquad  \frac{db}{dt}=\chi ac^{+}; \qquad  \frac{dc}{dt}=\chi ab^{+}
\end{equation}
that lead to the evolution of the pump mode number operator,
\begin{equation}
\frac{dN_{a}}{dt}=-\chi\left(ab^{+}c^{+}+a^{+}bc\right).
\end{equation}
It is easy to show the corresponding number operators for signal and idler modes are given by $dN_{b}/dt=dN_{c}/dt=-dN_{a}/dt$. To proceed, we will use the following  Manley-Rowe constants of motion\cite{manley:1956},
\begin{equation}\label{eq:MR}
M_{ab}=N_{a}+N_{b};\  \ M_{ac}=N_{a}+N_{c};\  \ M_{bc}=N_{b}-N_{c},
\end{equation}
expressing the underlying SU(2) and SU(1,1) symmetries in our model\cite{yurke:1986,brif:1996}. Differentiating Eqs.~(\ref{eq:semiQ_dt}) again results in decoupled equations of motion containing only commuting operators.
\begin{equation}\label{eq:number}\eqalign{
\frac{d^{2}N_{a}}{dt^{2}}=2\chi^{2}\left[3N_{a}^{2}-N_{a}\left(2M_{ab}+2M_{ac}+1\right)+M_{ab}M_{ac}\right]\\
\frac{d^{2}N_{b}}{dt^{2}}=-2\chi^{2}\left[3N_{b}^{2}-N_{b}\left(4M_{ab}-2M_{ac}-1\right)+M_{ab}\left(M_{ab}-M_{ac}-1\right)\right]\\
\frac{d^{2}N_{c}}{dt^{2}}=-2\chi^{2}\left[3N_{c}^{2}-N_{c}\left(4M_{ac}-2M_{ab}-1\right)+M_{ac}\left(M_{ac}-M_{ab}-1\right)\right]
}
\end{equation}
We now take the expectation value of the pump number operator equation~(\ref{eq:number}) for $N_a$ and make the approximation $\left<N_a^2\right>\approx\left<N_a\right>\left< N_a\right>$, the validity of which can be checked using full quantum numerical simulations of Eq.~(\ref{eq:semiQ_dt}) [see Sec.~\ref{sec:quantum}].  These approximations, along with the condition that both signal and idler mode start in the vacuum state, lead to the semiclassical evolution for the pump mode:
\begin{equation}\label{eq:semiNa}
N_{a}(\tau)=\beta_{+}+\left[N_{a}(0)-\beta_{+}\right]\mathrm{dn}\left[\sqrt{\beta_{+}-\beta_{-}}\tau,\frac{N_{a}(0)-\beta_{-}}{\beta_{+}-\beta_{-}}\right]^{-2},
\end{equation}
where $dn(u,k)$ is the Jacobi elliptic function and
\begin{equation}\label{eq:semibeta}
\beta_{\pm}=\frac{1}{4}\left[ 1+2N_{a}(0)\pm\sqrt{1+12N_{a}(0)+4N_{a}(0)^{2}}\right].
\end{equation}
It is important to note that both Eqs.~(\ref{eq:semiNa}) and (\ref{eq:semibeta}) are expressed in terms of the initial conditions of the pump mode only, a consequence of  the relations (\ref{eq:MR}) and the fact that the signal and idler are initially in the vacuum (ground) state. Equations of motion for both signal and idler can then be obtained by substitution of the c-number expressions for the pump mode with amplitude given by Eq.~(\ref{eq:semiNa}),
\begin{equation}\label{eq:cnum}\eqalign
a=\sqrt{N_{a}(\tau)}e^{-i\phi(t)}\\
a^{+}=\sqrt{N_{a}(\tau)}e^{i\phi(t)},
\end{equation}
where $\phi(t)$ is a slowly varying function of time, into Eq.~(\ref{eq:interaction}):
\begin{equation}
\tilde{H}_{I}=i\hbar\chi N_{a}(t)^{1/2}\left(b^{+}c^{+}e^{-i\phi(t)}-bc e^{i\phi(t)}\right).
\end{equation}
The time evolution of this now bilinear Hamiltonian can be straightforwardly calculated if we assume \cite{lu:1973}
\begin{equation}\label{eq:phase1}
\sqrt{N_{a}(t)}\frac{d\phi(t)}{dt}=\mathrm{const},
\end{equation}
which is the most general condition under which the Hamiltonian at different times commutes, $\left[\tilde{H}_{I}(t),\tilde{H}_{I}(t')\right]=0$.  Combining Eqs.~(\ref{eq:semiQ_dt}) and (\ref{eq:cnum}) results in the phase relation
\begin{equation}
a^{+}\frac{d a}{dt}-a\frac{d a^{+}}{dt}=-2i N_{a}\frac{d\phi(t)}{dt},
\end{equation}
which, expressing the left-hand side as Eq.~(\ref{eq:interaction}) through the Heisenberg equations for the operators Eq.~(\ref{eq:semiQ_dt}), yields a second condition for the pump phase:
\begin{equation}\label{eq:phase2}
2N_{a}(t)\frac{d\phi(t)}{dt}=\left<\tilde{H}_{I}\right>.
\end{equation}
As long as the initial state contains at least one mode in the vacuum state, we have the constant of motion $\left<H_{I}\right>=0$. Thus, Eqs.~(\ref{eq:phase1}) and (\ref{eq:phase2}) are only consistant if $\phi(t)=0$.  This leads to a bilinear Hamiltonian of the form,
\begin{equation}
\tilde{H}_{I}=i\hbar\chi N_{a}(t)^{1/2}\left(b^{+}c^{+}-bc\right).
\end{equation}
Using a similar analysis to that employed in Sec.~\ref{sec:parametric} gives,
\begin{equation}\eqalign{
\frac{d b(\tau)}{d\tau}=b(0)\cosh\left(\theta(\tau)\right)+c^{+}(0)\sinh\left(\theta(\tau)\right)\\
\frac{d c(\tau)}{d\tau}=c(0)\cosh\left(\theta(\tau)\right)+b^{+}(0)\sinh\left(\theta(\tau)\right),
}
\end{equation}
where,
\begin{equation}
\theta(\tau)=\int_{0}^{\tau}\sqrt{N_{a}\left(\tau'\right)}d\tau'.
\end{equation}
Therefore, we see that each mode behaves similarly as in the parametric approximation but with the number of particles dependent on the pump dynamics:
\begin{equation}
N_{b}(\tau)=N_{c}(\tau)=\sinh^{2}\left(\theta(\tau)\right).
\end{equation}
However, most importantly the modes are again in a thermal state with distribution
\begin{equation}
\left< O(
\tau)\right>=\mathrm{sech}^{2}\left(\sqrt{N_{a}(\tau)}\tau\right)\sum_{n}\tanh^{2n}\left(\sqrt{N_{a}(\tau)}\tau\right)\left<n| O| n\right>.
\end{equation}
Thus, the spectrum of the signal mode (Hawking radiation) remains thermal when pump mode energy loss due to signal and idler particle creation is taken into account within the semi-classical approximation.   

\section{Full Quantum Description}\label{sec:quantum}

\subsection{Short-time Approximation}\label{sec:short-time}
\begin{figure}[htbp]
\begin{center}
\includegraphics[width=5.5in]{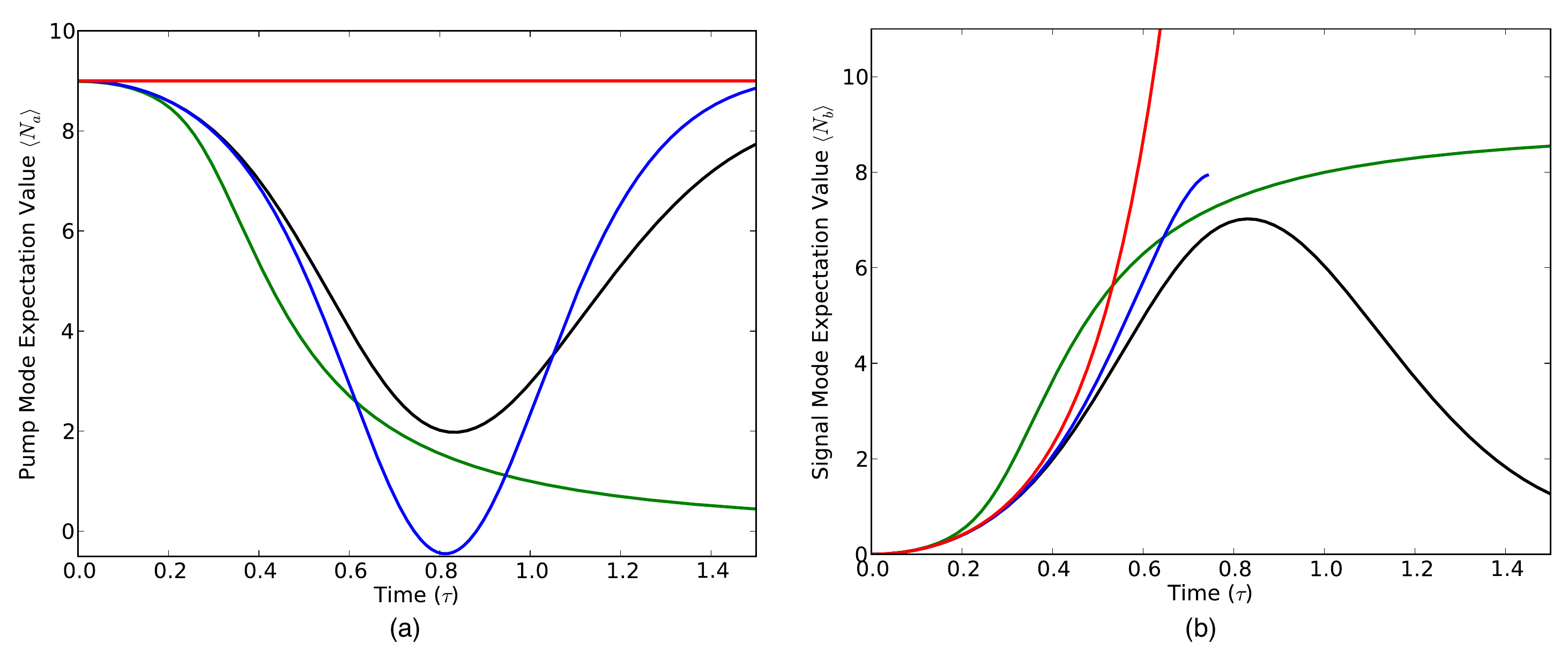}
\caption{a.)Evolution of pump mode initially in coherent state $\left< N_{a}(0)\right>=9$ for the parametric (red), semi-classical (blue),  short-time quantum approximation (green), and full quantum numerical solution (black), as a function of dimensionless time $\tau=\chi t$. b.)Corresponding population of the signal mode.  Note that the semi-classical analysis ceases to be valid when the pump mode becomes depleted.}
\label{fig:averages}
\end{center}
\end{figure}

In this section we consider the full quantum dynamics of the trilinear Hamiltonian Eq.~(\ref{eq:trilinear}), (i.e. quantize the pump mode as well).  Although the trilinear system does not allow for a general analytical solution, we can obtain approximate equations in the short time limit $\tau=\chi t\ll 1$.  Using the condition $\omega_{b}=\omega_{c}=\omega_{a}/2$, appropriate for modeling particle generation by a black hole, we can rewrite Eq.~(\ref{eq:trilinear}) ignoring constant terms as
\begin{equation}\label{eq:fullquantumH}
H=H_{0}+H_{\mathrm{int}}=\hbar\omega_{a}\left(a^{+}a+K_{z}\right)+i\hbar\chi\left(aK_{+}-a^{+}K_{-}\right),
\end{equation}
where
\begin{equation}
K_{+}=b^{+}c^{+},\ \ K_{-}=bc,\ \ K_{z}=\frac{1}{2}\left(b^{+}b+c^{+}c+1\right).
\end{equation}
We now construct the Casimir operator,
\begin{equation}
K^{2}=K_{z}^{2}-\frac{1}{2}\left(K_{+}K_{-}+K_{-}K_{+}\right)=\frac{1}{4}\left(M^{2}_{bc}-1\right),
\end{equation}
where the second equality comes from  definition (\ref{eq:MR}).  This operator obeys the eigenvalue equation
\begin{equation}
K^{2}\left|\Psi\right>=k(k-1) \left|\Psi\right>,
\end{equation}
where $k=1/2\left(\left| M_{bc}\right|+1\right)$.  We now look for simultaneous eigenstates of both $K^{2}$ and $K_{z}$: $\left| k,n_{c}\right>$, where $n_{c}$ is the number of idler particles.  These states can also be decomposed using the number state basis
\begin{equation}
\left| k,n_{c}\right>=\left| n_{c}+2k-1\right>_{b}\left| n_{c}\right>_{c}=\left| n_{b}\right>\left| n_{c}\right>,
\end{equation}  
with $n_{b}$ representing the number of particles in the signal.  Operating on this state with $K_{z}$ gives
\begin{equation}
K_{z}\left| k,n_{c}\right>=\left(k+n_{c}\right) \left| k,n_{c}\right>=\left(\frac{n_{b}+n_{c}+1}{2}\right)\left| n_{b}\right>\left| n_{c}\right>,
\end{equation}
valid only for initial conditions where both signal and idler modes are in pure states satisfying $n_{b}\ge n_{c}$.  Including the pump mode state in the full state vector and expressing the idler mode population as a function of the pump amplitude
\begin{equation}\label{eq:quantum-state}
\left| n_{a}\right>\left| k,M_{ac}-n_{a}\right>,
\end{equation}
one may switch to the interaction frame where the evolution depends only on $H_{\mathrm{int}}$ (\ref{eq:fullquantumH})
\begin{equation}\label{eq:quantum-evolution}
\left| \tau;k,M_{ac}\right>=e^{\tau\left(aK_{+}-a^{+}K_{-}\right)}\left|0;k,M_{ac}\right>,
\end{equation}
with the initial state with $M_{ac}=N_{a}(0)$ is given by
\begin{equation}\label{eq:initial}
\left| 0;k,M_{ac}\right>=\left| M_{ac}\right>\left|k,0\right>.
\end{equation}
The short time approximation $\tau=\chi t\ll1$ to Eq.~(\ref{eq:quantum-evolution}) can be calculated using the Baker-Campbell-Hausdorff formula truncated to $O\left(\tau^{2}\right)$
\begin{equation}\label{eq:truncate}
e^{\tau\left(aK_{+}-a^{+}K_{-}\right)}\approx e^{\tau aK_{+}}e^{-\tau a^{+}K_{-}}e^{-\frac{\tau^2}{2}\left[aK_{+},a^{+}K_{-}\right]}.
\end{equation}
Acting with the third exponential operator term on our initial state gives
\begin{equation}
e^{-\frac{\tau^2}{2}\left[aK_{+},a^{+}K_{-}\right]}\left| M_{ac}\right>\left|k,0\right>\approx \left(1-kM_{ac}\tau^{2}\right)\left| M_{ac}\right>\left|k,0\right>
\end{equation}
indicating clearly that the time over which the approximation is valid $t=\tau/\chi<1/(\chi\sqrt{kM_{ac}})$ decreases with pump amplitude and coupling strength.  Evaluating the BCH approximated expression~(\ref{eq:truncate}) on the initial state~(\ref{eq:initial}),  we obtain for the full evolution of the state:
\begin{equation}\label{eq:zstate}
\left|\tau;k,M_{ac}\right>=\frac{1}{\sqrt{N_{M_{ac}}(\tau)}}\sum_{n=0}^{M_{ac}}f_{n}\left(k,M_{ac}\right){\tau}^{n}\left|M_{ac}-n\right>\left|k,n\right>,
\end{equation}
where $N_{M_{ac}}(\tau)$ is the time-dependent normalization factor
\begin{equation}
N_{M_{ac}}(\tau)=e^{\tau^{-2}}\tau^{2M_{ac}}\Gamma\left(M_{ac}+1,\tau^{-2}\right),
\end{equation}
with $\Gamma\left(a,b\right)$ the reduced gamma function and
\begin{equation}
f_{n}\left(k,M_{ac}\right)=\left[\frac{M_{ac}!\Gamma\left(2k+n\right)}{n!\left(M_{ac}-n\right)!\Gamma\left(2k\right)}\right]^{1/2}.
\end{equation}
Our interest in the crossover from classical to quantum dynamics for the pump mode suggests that we use coherent states built from linear combinations of Eq.~(\ref{eq:zstate}) for the initial pump state.  To this end, we consider a general initial state
\begin{equation}\label{eq:initial-state}
\left|\Psi(0)\right>=\sum_{s=0}^{\infty}a_{s}\frac{f_{0}(s)}{\sqrt{N_{s}(0)}}\left|s\right>\left|0\right>\left|0\right>,
\end{equation}
with the pump mode in an as yet unspecified initial pure state with probability $P_{s}=\left|a_{s}\right|^2$ of being in state $s$ and both signal and idler modes in the vacuum state ($k=1/2$).  Here we implicitly assume the probabilities are normalized and add to unity.  The density matrix at some later time $\tau$ resulting from (\ref{eq:initial-state}) is given by
\begin{equation}
\rho_{abc}^{\Psi}(\tau)=\sum_{s=0}^{\infty}\sum_{r=0}^{\infty}\sum_{i=0}^{s}\sum_{j=0}^{r}a_{s}a_{r}^{*}\frac{f_{i}(s)\tau^{i}}{\sqrt{N_{s}(\tau)}}\frac{f_{j}(r)\tau^{j}}{\sqrt{N_{r}(\tau)}}\left|s-i\right>\left|i,i\right>\left<r-j\right|\left<j,j\right|,
\end{equation} 
which, after performing a partial trace, leads to the reduced density operators
\begin{equation}\label{eq:partiala}
\rho_{a}^{\Psi}(\tau)=\sum_{s=0}^{\infty}\sum_{r=0}^{\infty}\sum_{i=0}^{s}\sum_{j=0}^{r}a_{s}a_{r}^{*}\frac{f_{i}(s)\tau^{i}}{\sqrt{N_{s}(\tau)}}\frac{f_{j}(r)\tau^{j}}{\sqrt{N_{r}(\tau)}}\delta_{i,j}\left|s-i\right>\left<r-j\right|,
\end{equation}
where $\delta_{i,j}$ is the Kronecker delta, and
\begin{equation}\label{eq:partialb}
\rho_{b}^{\Psi}(\tau)=\sum_{s=0}^{\infty}\sum_{i=0}^{s}P_{s}\frac{f_{i}^{2}(s)}{N_{s}(\tau)}\tau^{2i}\left|i\right>\left<i\right|
\end{equation}
for the pump and signal modes respectively.  Fig.~\ref{fig:averages} compares the expectation values of the pump and signal modes for the parametric, semi-classical, and short-time quantum approximations, as well as the full numerical solution to Eqs.~(\ref{eq:semiQ_dt}) where the pump mode is initially in a coherent state with amplitude $\left<N_{a}(0)\right>=9$ corresponding to classical pump modes with amplitude $A=3$, and the signal and idler modes are initially in their vacuum (ground) states.

\subsection{Mode Spectrum Dynamics Under the Short-time Approximation}\label{sec:short-dynamics}
The main limitation of the short-time approximation is the inability to account for backreaction effects resulting from the build up of quanta in the signal and idler modes.  For particles produced in the Hawking process, however,  the entangled pairs generated early in the evolution do not, to first-order, effect those created later from the black hole\cite{mathur:2009}. Furthermore, the emitted radiation does not build up in the vicinity of the black hole, but  escapes to spatial infinity. Therefore, in this section we will suppose that the expressions derived above for the evolving pump and signal states, Eqs.~(\ref{eq:partiala}) and (\ref{eq:partialb}) respectively, in fact provide a more relevant zero-dimensional model of a radiating black-hole when extrapolated beyond their original short-time domain of validity.   The pump and signal modes contain in general a large number of coefficients for each number state and cannot be easily evaluated.  However, in the long-time limit where the pump is nearly depleted, these equations can be considerably simplified by noting that a general element of Eq.~(\ref{eq:partialb}) for fixed $s$ is given by
\begin{equation}
P_{s}\frac{f_{i}^{2}(s)}{N_{s}(\tau)}\tau^{2i}=P_{s}\frac{e^{-\tau^{-2}}}{u!\tau^{2u}}\frac{\Gamma(s+1)}{\Gamma\left(s+1,\tau^{-2}\right)},\ \ u=s-i.
\end{equation}
As such, only those elements where $i=s$ remain nonzero, leading to the signal mode density matrix
\begin{equation}\label{eq:long-time-state}
\rho^{\Psi}_{b}=\sum_{s=0}^{\infty}P_{s}\left|s\right>\left<s\right|,
\end{equation}
which is in general a mixed state with number-state probability distribution $P_{s}$ identical to that of the initial pump state.  Therefore, by measuring the late-time signal or idler mode distribution we recover all but the phase information associated with the initial pure state of the pump mode.  Additionally, Eq.~(\ref{eq:long-time-state}) shows that the number of quanta in the signal mode is equal to the initial number of quanta in the pump mode, indicating that the pump (\ref{eq:partiala}) ends up in the vacuum (ground) state.  

\begin{figure}[htbp]
\begin{center}
\includegraphics[width=5.5in]{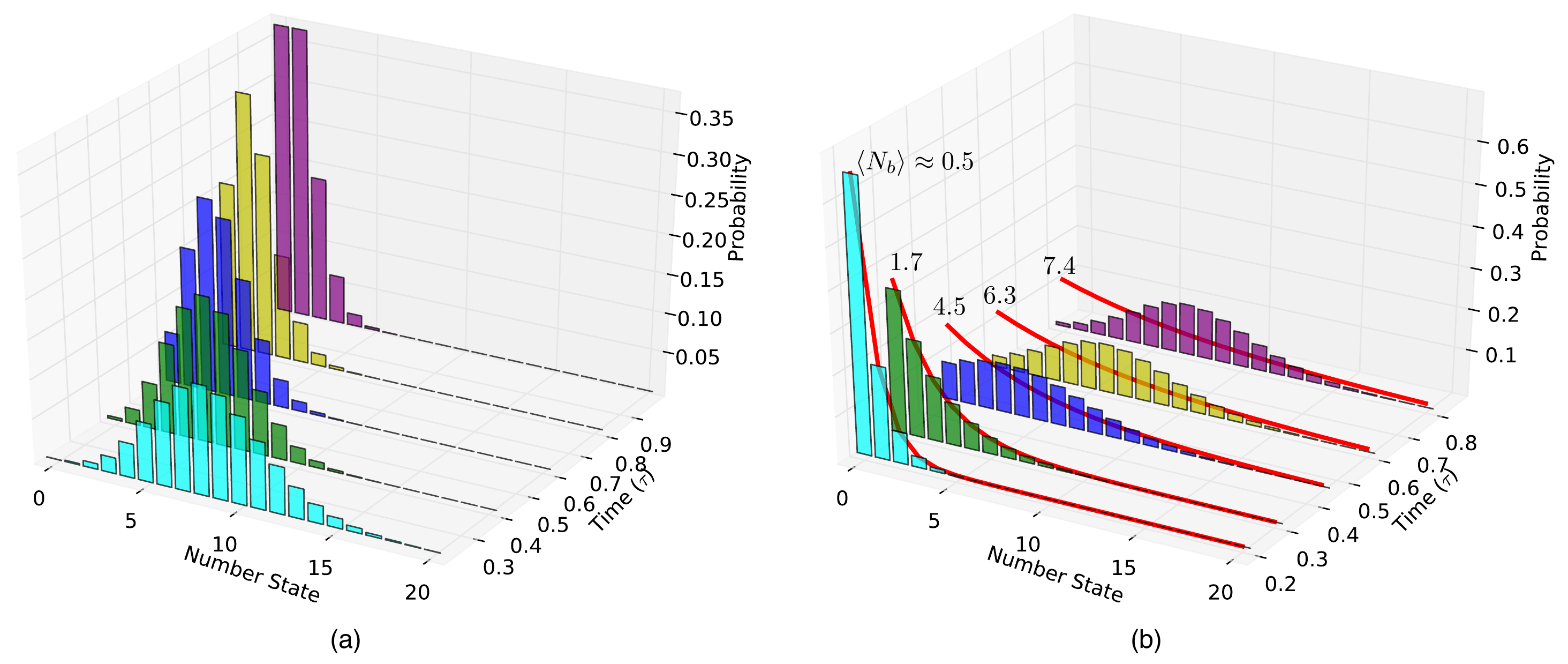}
\caption{a) Spectrum of the pump mode at several time-steps when the pump is initially in a coherent state with $\left<N_{a}(0)\right>=9$. b) Corresponding spectrum of the signal mode.  Red lines indicate what the distribution would be at each time-step if in a thermal state corresponding to occupation number $\left<N_{b}(\tau)\right>$, Eq.~(\ref{eq:bose}).}
\label{fig:spectrum}
\end{center}
\end{figure}

The most important knowledge gained from Eq.~(\ref{eq:long-time-state}) is that the signal mode spectrum will no longer be that of a thermal state, in contrast to the parametric and semi-classical approximations in Sec.~\ref{sec:parametric} and Sec.~\ref{sec:semiclassical}, respectively.  Focusing on coherent states, in Fig.~\ref{fig:spectrum} we give an example of the evolution of Eqs.~(\ref{eq:partiala}) and (\ref{eq:partialb}) by plotting the probability distributions for both modes as a function of time $\tau$ for a pump initially in a coherent state with amplitude $\left<N_{a}(0)\right>=9$ and initial vacuum state for both signal and idler.
Additionally, in Fig.~\ref{fig:spectrum}b we highlight what the thermal distribution would be at each time step by equating $\left<N_{b}(\tau)\right>$ with a Bose-Einstein distribution to extract an effective temperature:
\begin{equation}\label{eq:bose}
\left<N_{b}(\tau)\right>=\left[e^{\hbar\omega_{b}/k_{\mathrm{b}}T_{\mathrm{eff}}\left(\tau\right)}-1\right]^{-1}.
\end{equation}
In Fig.~\ref{fig:spectrum}a see we the evolution of the initial coherent state as it loses quanta to the signal and idler modes and evolves towards the ground state represented by a peak in the probability distribution at the $n=0$ number state value.  The corresponding evolution of the signal mode in Fig.~\ref{fig:spectrum}b starts with the vacuum state and progresses towards the state with probability distribution identical to that of the initial pump coherent state, Eq.~(\ref{eq:long-time-state}).  By comparison with the effective thermal state~(\ref{eq:bose}), we can see that the initial probability distribution for the signal mode is nearly that of a thermal state up until the pump mode has transferred nearly half of its initial energy corresponding to $\left<N_{b}(\tau)\right>=4.5$.  As the evolution continues, Fig.~\ref{fig:spectrum}b shows the increasing deviation from the effective thermal description for the signal mode distribution as expected from Eq.~(\ref{eq:long-time-state}).

\subsection{Non-thermal Spectra and Information}\label{sec:info}
\begin{figure}[htbp]
\begin{center}
\includegraphics[width=5.5in]{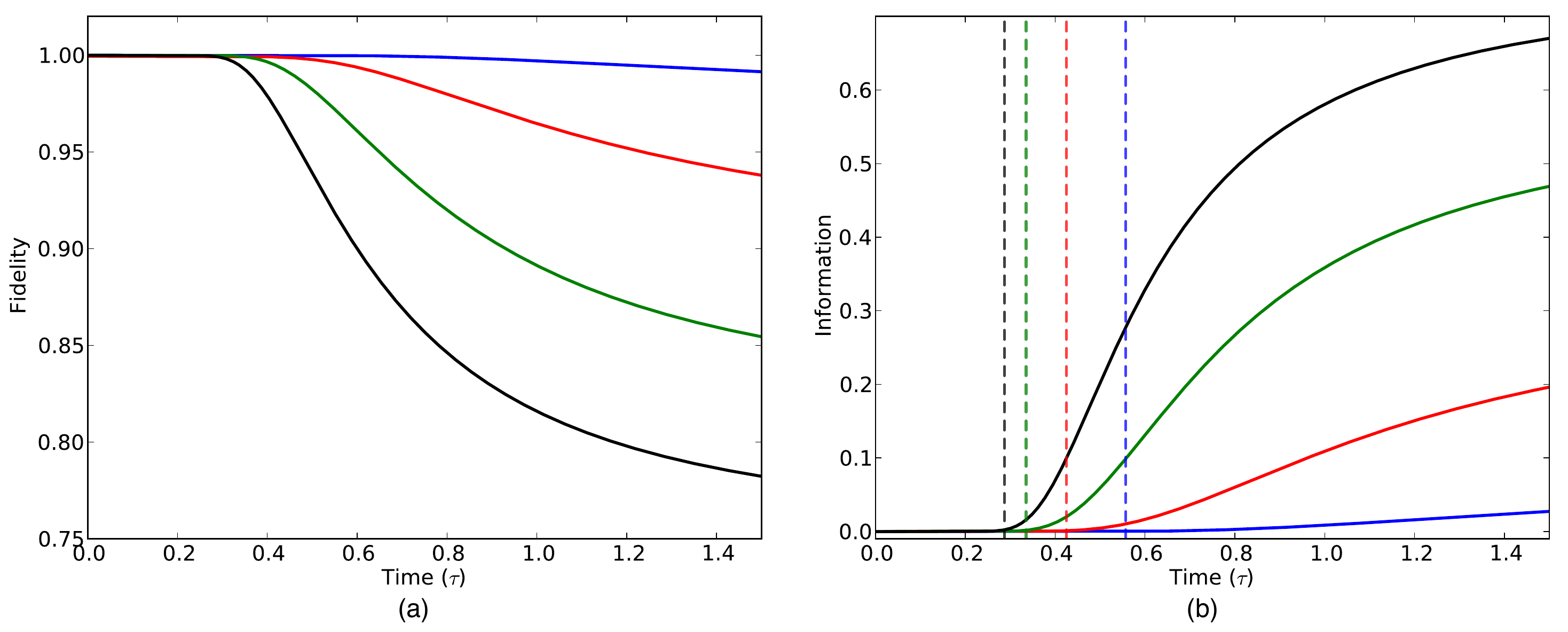}
\caption{a) Fidelities of the signal mode given a pump mode initially in a coherent state with occupation number $\left<N_{a}\right>=1$ (blue), $3$ (red), $6$ (green), and $9$ (black) under the short-time approximation.  b)  Information present in the signal spectrum calculated from Eq.~(\ref{eq:information}).  Equivalently colored vertical lines denote the time at which the effective subspace dimensions satisfy, $d^{\mathrm{eff}}_{a}=d^{\mathrm{eff}}_{bc}$.}
\label{fig:fig4}
\end{center}
\end{figure}

We now quantify the deviations of the signal mode spectrum Eq.~(\ref{eq:partialb}) from that of a thermal state using the fidelity\cite{nielson:2000}
\begin{equation}\label{eq:fidelity}
F_{b}(\tau)=\mathrm{Tr}\sqrt{\rho_{b}(\tau)^{1/2}\sigma(\tau)\rho_{b}(\tau)^{1/2}},
\end{equation}
where $\rho_{b}(\tau)$ is the density matrix of the signal mode and $\sigma(\tau)$ is a thermal density matrix with effective temperature determined by Eq.~(\ref{eq:bose}) using the occupation number $\left<N_{b}\left(\tau\right)\right>$.  The fidelity provides a measure of the distance between these two states in distribution space with range $0\le F\le 1$, where unity indicates that the two density matrices are identical.  In Fig.~\ref{fig:fig4}a we plot the fidelity of the signal mode assuming a pump mode initially in a coherent state starting with several occupation numbers ranging from one to nine.  By comparison with Fig.~\ref{fig:averages}, we can see that for the initial coherent state $\left<N_{a}(0)\right>=9$ the fidelity remains essentially unity, indicating that the signal mode density matrix Eq.~(\ref{eq:partialb}) is nearly thermal, until the pump has transferred close to half of its initial energy into the signal and idler modes beyond which point there is a strong deviation from the thermal state as the signal mode asymptotically approaches the state given by Eq.~(\ref{eq:long-time-state}).  This is in agreement with the qualitative description presented in Fig.~\ref{fig:spectrum}b and remains true for all the initial states considered in Fig.~\ref{fig:fig4}a. 

This deviation from a thermal distribution also indicates that the signal mode spectrum contains information defined as\cite{page:1993}
\begin{equation}\label{eq:information}
I_{b}(\tau)=S_{b}^{\mathrm{th}}(\tau)-S_{b}(\tau),
\end{equation}
where $S_{b}^{\mathrm{th}}(\tau)$ is the Von Neumann entropy of the signal mode in a thermal state with equal average occupation number, Eq.~(\ref{eq:thermal-entropy}), and $S_{b}(\tau)$ is the actual entropy of the mode calculated using Eq.~(\ref{eq:VNentropy}) and the reduced density matrix of the signal mode $\rho_{b}(\tau)$.  In Fig.~4b we plot the information contained in the signal mode for the initial pump coherent states considered in Fig.~4a.  Given that the fidelity is nearly unity up until the pump mode transfers half  of its original quanta to the signal and idler modes, it follows that the information content of the signal or idler mode is close to zero over this same time interval before increasing as the signal spectrum becomes identical to that of the initial coherent states.  

In order to account for the dynamics of the signal mode information, we first consider a general bipartite pure state of a system with fixed total energy that is composed of subsystems $A$ and $B$, each with finite Hilbert space dimensions $d_{A}$ and $d_{B}$, respectively.  It is known that subsystem $B$ will be nearly thermal and thus contain approximately no information as long as $d_{A}\gg d_{B}$, with the information content of subsystem $B$ becoming apparent only after $d_{A}\approx d_{B}$\cite{page:1993,page:1993-2,popescu:2006}.  Similar dynamics for the information content of the signal mode is shown in Fig.~\ref{fig:fig4}. However, the three harmonic oscillator modes from which a pure state of the trilinear Hamiltonian Eq.~(\ref{eq:trilinear}) is composed all contain an infinite set of states, preventing the direct application of dimensional arguments to our model.  One can, however, define an effective subspace dimension for each mode\cite{popescu:2006}
\begin{equation}\label{eq:effective}
d^{\mathrm{eff}}_{i}(\tau)=\frac{1}{\mathrm{Tr}\left[\sigma_{i}^{2}(\tau)\right]},\ \ i=a,b,c
\end{equation}
determined by the purity of the effective thermal state $\sigma_{i}(\tau)$ with temperature given by Eq.~(\ref{eq:bose}).  This definition is motived by the fact that the purity of a mixed state density matrix is proportional to the number of basis states over which the fractional population of the mixed state is nonzero.  For a state with an energy (quanta) constraint, using the thermal state $\sigma(\tau)$ gives the minimal value for the purity, and as such Eq.~(\ref{eq:effective}) represents an effective maximum number of states constrained by the number of quanta in the mode at time $\tau$.  Therefore, if we partition our initial pure state into bipartite subsystems $d_{a}$ and $d_{bc}$ composed of pump and combined signal and idler modes respectively, then the initial effective subspace dimensions satisfy $d^{\mathrm{eff}}_{a}\gg d^{\mathrm{eff}}_{bc}=\left(d^{\mathrm{eff}}_{b}\right)^{2}$, where the equality is due to the symmetry between signal and idler modes.  In Fig.~\ref{fig:fig4}b we plot the times at which $d^{\mathrm{eff}}_{a}=d^{\mathrm{eff}}_{bc}$ for each of the initial pump mode coherent states considered in Fig.~\ref{fig:fig4}a.  Just as for finite dimensional pure states, the information contained in the signal or idler subsystems remains nearly zero until after $d^{\mathrm{eff}}_{a}(\tau)\approx d^{\mathrm{eff}}_{bc}(\tau)$, provided we define the effective subspace dimension using Eq.~(\ref{eq:effective}).  Our results are in agreement with those of Page\cite{page:1993} where a similar argument was put forth for the evolution of information in the Hawking radiation from a finite dimensional black hole.

\subsection{Numerical Results for the Full Trilinear Evolution}
\begin{figure}[htbp]
\begin{center}
\includegraphics[width=5.5in]{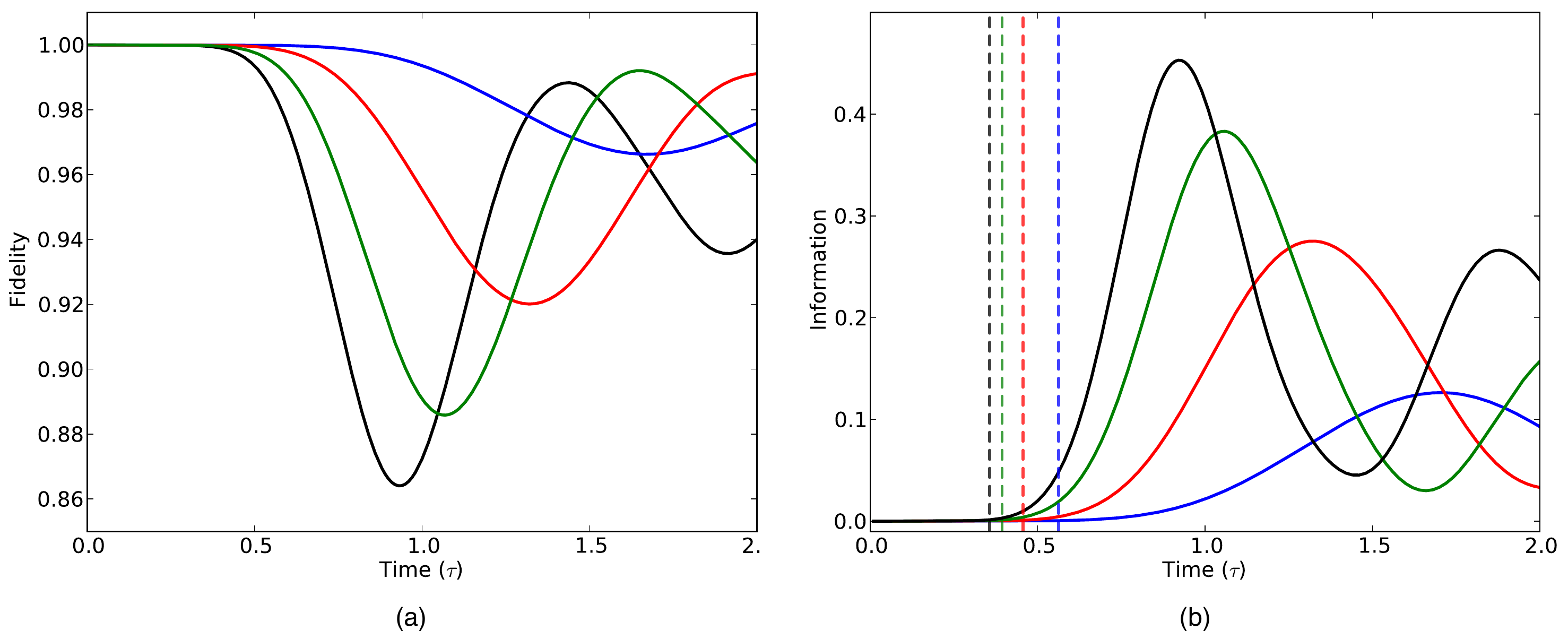}
\caption{a) Fidelities of the signal mode using numerical simulations of Eqs.~(\ref{eq:semiQ_dt}) with a pump mode initially in a coherent state with occupation number $\left<N_{a}\right>=1$ (blue), $3$ (red), $6$ (green), and $9$ (black).  b)  Corresponding information present in the signal mode spectrum and times at which $d^{\mathrm{eff}}_{a}(\tau)=d^{\mathrm{eff}}_{bc}(\tau)$ (dashed lines).}
\label{fig:fig5}
\end{center}
\end{figure}

Unlike Hawking radiation, which is well-modeled using the quantum short-time approximation, backaction due to the build up of quanta in the signal and idler modes has a strong effect on the evolution of the trilinear Hamiltonian (\ref{eq:interaction}).  In Fig.~\ref{fig:averages} we see that backreaction effects quickly lead to differing evolutions between the short-time and full quantum dynamics; backreaction prevents the full transfer of quanta from the pump mode\cite{drobny:1994}, and results in the oscillation of quanta between pump and signal/idler modes. To account for these backaction terms we repeat the analysis in Sec.~\ref{sec:info} by numerically calculating the full dynamics of Eq.~(\ref{eq:semiQ_dt}).  Fig.~\ref{fig:fig5} shows the modifications to both the fidelity and information content of the signal mode when backaction is included in the dynamics, as well as the subspace dimension condition $d^{\mathrm{eff}}_{a}(\tau)=d^{\mathrm{eff}}_{bc}(\tau)$.  
As expected, the short-time dynamics in Fig.~\ref{fig:fig5} are nearly identical to those in Fig.~\ref{fig:fig4} up until $d^{\mathrm{eff}}_{a}(\tau)\approx d^{\mathrm{eff}}_{bc}(\tau)$ which occurs later than in the short-time approximation as the backaction from particles in the signal mode begins to impede the transfer of quanta from the pump.  The peaks in the information content of the signal mode correspond to the times at which backaction from the pump mode has completely stopped the transfer of energy between modes.  With the majority of quanta now in the signal and idler modes, the flow of energy reverses directions as the signal and idler now drive the pump mode leading to the oscillations seen in Fig.~\ref{fig:fig5}. As a model for black hole evaporation these oscillations represent the unphysical process of Hawking radiation flowing back into the black hole; the connection between the trilinear Hamiltonian and Hawking emission is valid only for the initial transfer of quanta from pump to signal/idler modes.

\section{Tripartite Entanglement}\label{sec:entanglement}
\begin{figure}[htbp]
\begin{center}
\includegraphics[width=5.5in]{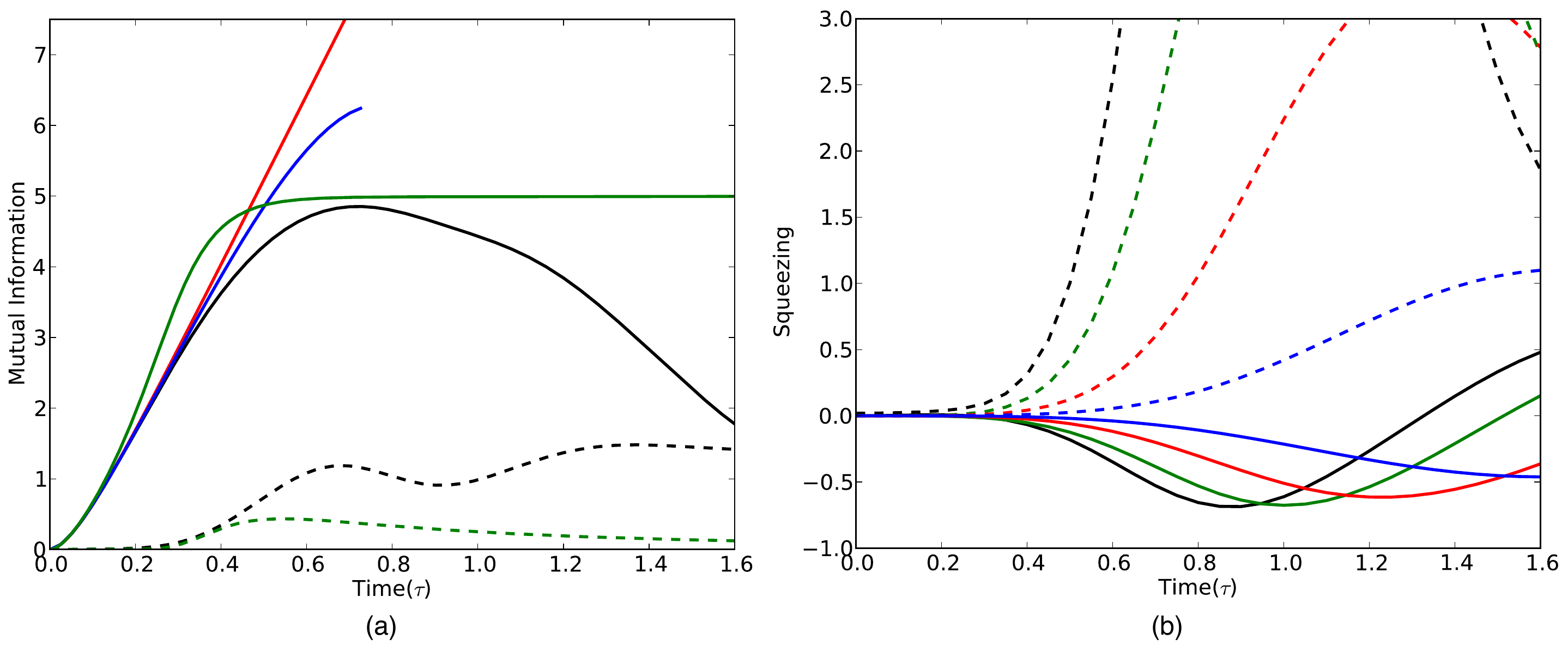}
\caption{a) Signal-idler mutual information $I_{b-c}$ for pump with $\left<N_{a}(0)\right>=9$ in the parametric (red), semi-classical (blue), quantum short-time (green), and full numerical solution (black).  Dashed lines represent mutual information $I_{a-bc}$ indicating entanglement with the pump mode.  b) Quadrature squeezing parameters $q_{\pm}$ of the pump mode given the pump is initially in a coherent state with occupation number $\left<N_{a}\right>=1$ (blue), $3$ (red), $6$ (green), and $9$ (black).  Negative values indicating squeezing of the $q_{-}$ quadrature.  Dashed lines represent the corresponding $q_{+}$ quadrature.}
\label{fig:fig6}
\end{center}
\end{figure}

With the pump mode quantized, one can consider the entanglement between pump and signal/idler modes.  In Sec.~\ref{sec:parametric}, it was shown that ignoring this entanglement (pump mode treated classically) and tracing over one mode of a two-mode squeezed state leads directly to the thermal characteristics of the remaining system.  In the full quantum description the statistics of the signal mode is obtained by tracing over both idler and pump modes and as such the distribution of entanglement between modes of this now tripartite system is important for characterizing the spectrum of the signal mode.  Additionally, entanglement with the pump mode may appreciably alter the state dynamics of the pump compared to the classical approximations in Secs.~\ref{sec:parametric} and \ref{sec:semiclassical}.  Given that multipartite entanglement is not as well understood as for the bipartite case, we begin by again considering the system to be partitioned into two subsystems consisting of the pump mode and the combined signal-idler.  This bipartite separation into pump and signal-idler subsystems allows us to use the mutual information\cite{nielson:2000}
\begin{equation}\label{eq:mutual}
I_{a-bc}=S_{a}+S_{bc}-S_{abc}
\end{equation}
as a measure of total correlations, both classical and quantum, between subsystems\cite{groisman:2005}.  Since our tripartite system state remains pure throughout its evolution, the total entropy $S_{abc}=0$ and the subsystem entropies satisfy $S_{a}=S_{bc}$.  The mutual information is therefore twice the entropy of the pump mode subsystem, $I_{a-bc}=2S_{a}$.  Thus, the subsystem entropy of the pump is a direct measure of entanglement with the signal and idler modes, which is not taken into account in either the parametric or semi-classical solutions.  Of course the signal and idler subsystems are also entangled with each other as in Secs.~\ref{sec:parametric} and \ref{sec:semiclassical}.  With both signal and pump modes in identical states, the signal-idler mutual information is given by
\begin{equation}\label{eq:bc-info}
I_{b-c}=S_{b}+S_{c}-S_{bc}=2S_{b}-S_{a}
\end{equation}
indicating the tradeoff between the entanglement of the signal/idler and that of the pump mode Eq.~(\ref{eq:mutual}). In Fig.~\ref{fig:fig6}a we plot the mutual information Eqs.~(\ref{eq:mutual}) and (\ref{eq:bc-info}) for a pump initially in a coherent state of amplitude $\left<N_{a}(0)\right>=9$ for the parametric, semi-classical and short-time approximations along with the full numerical solution obtained using Eqs.~(\ref{eq:semiQ_dt}).  We see the mutual information (pump entanglement) $I_{a-bc}$ begins to become appreciable around the same time as the information content of the signal mode becomes apparent in Figs.~\ref{fig:fig4}b and \ref{fig:fig5}b indicating the increasing role of the pump mode in the dynamics.  The numerical solution shows that this increase in pump entanglement reduces the signal-idler mutual information $I_{b-c}$ with respect to the semi-classical and parametric approximations as expected from Eq.~(\ref{eq:bc-info}).  In the full quantum dynamics the pump is never allowed to be fully depleted and is therefore always entangled with the signal and idler.  However in the short-time approximation the late-time pump mode is depleted and approaches the ground state where $S_{a}=0$.

Finally, we show that the entanglement buildup with the pump mode given by $I_{a-bc}$ not only affects the signal and idler modes, but also results in amplitude dependent squeezing of the pump.  In Fig.~\ref{fig:fig6}b we plot the amplitude of the squeezing parameters\cite{walls:2008},
\begin{equation}
q_{\pm}=4\left<\Delta X_{\pm}^{2}\right>-1, \ \ X_{\pm}=\frac{a(\tau)\pm a^{+}(\tau)}{2}
\end{equation}
 for the pump mode, where $\left<\Delta X_{\pm}^{2}\right>$ is the variance of the quadratures.  As can be seen, the entanglement with the signal and idler modes leaves the pump mode in a non-classical squeezed state over the time range of interest.

\section{Conclusion}\label{sec:conclusion}
In this paper we have investigated the effect of a dynamical quantized pump mode on the generation of quanta in a parametric amplifier.  We have shown that a quantized pump mode leads naturally to a non-thermal spectrum for the signal and idler modes once the pump has released nearly half of its initial energy, such that the effective subspace dimensions of the pump and signal/idler mode systems approximately coincide.  The departure from a thermal state indicates that the signal spectrum contains information that may be used to partially reconstruct the initial state of the pump mode. Once quantized, the pump mode becomes entangled with the signal and idler leading to non-classical squeezed states of the pump. As a simple, zero-dimensional model of the Hawking effect, the present findings lend support to  the possibility for non-thermal emission from a quantum (as opposed to semiclassical) black hole; the  emitted radiation is entangled with the quantized gravitational degrees of freedom and yields information concerning initial formation of the black hole.

\ack
This work was partially supported by the JSPS Summer Program and the National Science Foundation under Grant No's. OISE-0913132 (P.D.N), CMS-0404031 and DMR-0804477 (M.P.B).  P.D.N thanks W.G. Brown for helpful discussions as well as H. Yamaguchi and the NTT Basic Research Laboratories for their hospitality and support where part of this work was carried out.

\bibliographystyle{unsrt}
\bibliography{trilinear}

\begin{thebibliography}{10}

\bibitem{hawking:1974}
S.~W. Hawking.
\newblock Black hole explosions.
\newblock {\em Nature}, 248:30, 1974.

\bibitem{hawking:1975}
S.~W. Hawking.
\newblock Particle creation by black holes.
\newblock {\em Commun. Math. Phys.}, 43:199, 1975.

\bibitem{hartle:1976}
J.~B. Hartle and S.~W. Hawking.
\newblock Path-integral derivation of black-hole radiance.
\newblock {\em Phys. Rev. D}, 13:2188, 1976.

\bibitem{boulware:1976}
D.~G. Boulware.
\newblock Hawking radiation and thin shells.
\newblock {\em Phys. Rev. D}, 13:2169, 1976.

\bibitem{gibbons:1977}
G.~W. Gibbons and S.~W. Hawking.
\newblock Action integrals and partition functions in quantum gravity.
\newblock {\em Phys. Rev. D}, 15:2752, 1977.

\bibitem{parentani:2000}
R.~Parentani.
\newblock Hawking radiation from feynman diagrams.
\newblock {\em Phys. Rev. D}, 61:27501, 2000.

\bibitem{bombelli:1986}
L.~Bombelli, R.~K. Koul, J.~Lee, and R.~D. Sorkin.
\newblock A quantum source of entropy for black holes.
\newblock {\em Phys. Rev. D}, 34:373, 1986.

\bibitem{srednicki:1993}
M.~Srednicki.
\newblock Entropy and area.
\newblock {\em Phys. Rev. Lett.}, 71(666), 1993.

\bibitem{eisert:2010}
J.~Eisert, M.~Cramer, and M.~B. Plenio.
\newblock Colloquium: Area laws for the entanglement entropy.
\newblock {\em Rev. Mod. Phys.}, 82:277--306, 2010.

\bibitem{aharanov:1987}
Y.~Aharanov, A.~Casher, and S.~Nussinov.
\newblock The unitary puzzle and planck mass stable particles.
\newblock {\em Phys. Lett. B}, 191:51, 1987.

\bibitem{giddings:1992}
S.~B. Giddings.
\newblock Black holes and massive remnants.
\newblock {\em Phys. Rev. D}, page 1347, 46.

\bibitem{hawking:1988}
S.~W. Hawking.
\newblock Wormholes in spacetime.
\newblock {\em Phys. Rev. D}, 37(4):904, 1988.

\bibitem{frolov:1990}
V.~P. Frolov, M.~A. Markov, and V.~F. Mukhanov.
\newblock Black holes as possible sources of closed and semiclosed worlds.
\newblock {\em Phys. Rev. D}, 41:383, 1990.

\bibitem{page:1993}
D.~N. Page.
\newblock Information in black hole radiation.
\newblock {\em Phys. Rev. Lett.}, 71:3743, 1993.

\bibitem{parikh:2000}
M.~K. Parikh and F.~Wilczek.
\newblock Hawking radiation as tunneling.
\newblock {\em Phys. Rev. Lett.}, 85:5042, 2000.

\bibitem{hawking:2005}
S.~W. Hawking.
\newblock Information loss in black holes.
\newblock {\em Phys. Rev. D}, 72:084013, 2005.

\bibitem{terno:2005}
D.~R. Terno.
\newblock From qubits to black holes: Entropy, entanglement and all that.
\newblock {\em Int. J. Mod. Phys. D}, 14:2307, 2005.

\bibitem{jacobson:1995}
T.~A. Jacobson.
\newblock Thermodynamics of spacetime: The einstein equation of state.
\newblock {\em Phys. Rev. Lett.}, 75:1260, 1995.

\bibitem{carlip:2008}
S.~Carlip.
\newblock Is quantum gravity necessary?
\newblock {\em Class. Quant. Grav.}, 25:154010, 2008.

\bibitem{dicke:1953}
R.~H. Dicke.
\newblock Coherence in spontaneous radiation processes.
\newblock {\em Phys. Rev.}, 93:99, 1953.

\bibitem{mollow:1967}
B.~R. Mollow and R.~J. Glauber.
\newblock Quantum theory of parametric amplification. i.
\newblock {\em Phys. Rev.}, 160:1076, 1967.

\bibitem{travis:1968}
M.~Travis and F.~W. Cummings.
\newblock Exact solutions for an n-molecule-radiation-field hamiltonian.
\newblock {\em Phys. Rev.}, 170:379, 1968.

\bibitem{tucker:1969}
J.~Tucker and D.~F. Walls.
\newblock Quantum theory of the traveling-wave frequency converter.
\newblock {\em Phys. Rev.}, 178:2036, 1969.

\bibitem{walls:1970}
D.~F. Walls and R.~Barakat.
\newblock Quantum-mechanical amplification and frequency coonversion with a
  trilinear hamiltonian.
\newblock {\em Phys. Rev. A}, 1:446, 1970.

\bibitem{lu:1973}
E.~Y.~C. Lu.
\newblock Quantum theory of nonlinear optical processes with time-dependent
  pump amplitude and phase: Frequnecy conversion.
\newblock {\em Phys. Rev. A}, 8:1053, 1973.

\bibitem{agrawal:1974}
G.~P. Agrawal and C.~L. Mehta.
\newblock Dynamics of parametric processes with a trilinear hamiltonian.
\newblock {\em J. Phys. A}, 7:607, 1974.

\bibitem{mcneil:1983}
K.~J. McNeil and C.~W. Gardiner.
\newblock Quantum statistics of parametric oscillation.
\newblock {\em Phys. Rev. A}, 28:1560, 1983.

\bibitem{gerlach:1976}
U.~H. Gerlach.
\newblock The mechanism of blackbody radiation from an incipient black hole.
\newblock {\em Phys. Rev. D}, 14(6):1479, 1976.

\bibitem{barnett:1985}
S.~M. Barnett and P.~L. Knight.
\newblock Thermofield analysis of squeezing and statistical mixtures in quantum
  optics.
\newblock {\em J. Opt. Soc. Am. B}, 2(3):467, 1985.

\bibitem{yurke:1987}
B.~Yurke and M.~Potasek.
\newblock Obtainment of thermal noise from a pure state.
\newblock {\em Phys. Rev. A}, 36(7):3464, 1987.

\bibitem{boyd:2003}
R.~W. Boyd.
\newblock {\em Nonlinear Optics}.
\newblock Academic Press, 2nd edition, 2003.

\bibitem{truax:1985}
D.~R. Traux.
\newblock Baker-campbell-hausdorff relations and unitarity of su(2) and su(1,1)
  squeeze operators.
\newblock {\em Phys. Rev. D}, 31:1988, 1985.

\bibitem{fuentes:2005}
I.~Fuentes-Schuller and R.~B. Mann.
\newblock Alice falls into a black hole: Entanglement in noninertial frames.
\newblock {\em Phys. Rev. Lett.}, 95(120404), 2005.

\bibitem{walls:2008}
D.~F. Walls and G.~J. Milburn.
\newblock {\em Quantum Optics, 2nd Ed.}
\newblock Springer, 2008.

\bibitem{kibble:1980}
T.~W.~B. Kibble and S.~Randjbar-Daemi.
\newblock Non-linear coupling of quantum theory and classical gravity.
\newblock {\em J. Phys. A}, 13:141, 1980.

\bibitem{horowitz:1980}
G.~T. Horowitz.
\newblock Semiclassical relativity: The weak-field limit.
\newblock {\em Phys. Rev. D}, 21:1445, 1980.

\bibitem{anselmi:2007}
D.~Anselmi.
\newblock Renormalization and causality violations in classical gravity coupled
  with quantum matter.
\newblock {\em JHEP}, 1:62, 2007.

\bibitem{unruh:1984}
W.~G. Unruh.
\newblock Steps toward a quantum theory of gravity.
\newblock In {\em Quantum Theory of Gravity: Essays in Honor of the 60th
  Birthday if Bryce S. De Witt}. Bristol: Hilger, 1984.

\bibitem{manley:1956}
J.~M. Manley and H.~E. Rowe.
\newblock Some general properties of nonlinear elements-part i. general energy
  relations.
\newblock {\em Proc. of IRE}, 44:904, 1956.

\bibitem{yurke:1986}
B.~Yurke, S.~L. McCall, and J.~R. Klauder.
\newblock su(2) and su(1,1) interferometers.
\newblock {\em Phys. Rev. A}, 33:4033, 1986.

\bibitem{brif:1996}
C.~Brif.
\newblock Coherent states for quantum systems with a trilinear boson
  hamiltonian.
\newblock {\em Phys. Rev. A}, 54:5253, 1996.

\bibitem{nielson:2000}
M.~A. Nielson and I.~L. Chuang.
\newblock {\em Quantum Computation and Quantum Information}.
\newblock Cambridge University Press, 2000.

\bibitem{page:1993-2}
D.~N. Page.
\newblock Average entropy of a subsystem.
\newblock {\em Phys. Rev. Lett.}, 71(9):1291, 1993.

\bibitem{popescu:2006}
S.~Popescu, A.~J. Short, and A.~Winter.
\newblock Entanglement and the foundations of statistical mechanics.
\newblock {\em Nature Phys.}, 2:754, 2006.

\bibitem{mathur:2009}
S.~D. Mathur.
\newblock The information paradox: a pedagogical introduction.
\newblock {\em Class. Quant. Grav.}, 26:224001, 2009.

\bibitem{drobny:1994}
G.~Drobn\'y and V.~Bu\v{z}ek.
\newblock Fundamental limit on energy transfer in k-photon down-conversion.
\newblock {\em Phys. Rev. A}, 50(4):3492, 1994.

\bibitem{groisman:2005}
B.~Groisman, S.~Popescu, and A.~Winter.
\newblock Quantum, classical, and total amount of correlations in a quantum
  system.
\newblock {\em Phys. Rev. A}, page 032317, 2005.

\end{thebibliography}
\end{document}